\documentstyle[12pt]{article}
\begin{document}
\def\beq{\begin{equation}}
\def\eeq{\end{equation}}
\def\bea{\begin{eqnarray}}
\def\eea{\end{eqnarray}}
\def\ve{\vert}
\def\vel{\left|}
\def\ver{\right|}
\def\nnb{\nonumber}
\def\ga{\left(}
\def\dr{\right)}
\def\aga{\left\{}
\def\adr{\right\}}
\def\rar{\rightarrow}
\def\nnb{\nonumber}
\def\la{\langle}
\def\ra{\rangle}
\def\ba{\begin{array}}
\def\ea{\end{array}}
\def\tep{$B \rar K \ell^+ \ell^-$}
\def\tepm{$B \rar K \mu^+ \mu^-$}
\def\tept{$B \rar K \tau^+ \tau^-$}

\title{Light cone QCD sum rule analysis
            of $B\to K\ell^+\ell^-$ decay }

\author{ {\small T. M. AL\.{I}EV$^1$ \thanks
{e-mail:taliev@rorqual.cc.metu.edu.tr}\,\,,
H. KORU$^2$\,\,,
A. \"{O}ZP\.{I}NEC\.{I}$^1$ \, and \,
M. SAVCI$^1$ }\\
{\small 1) Physics Department, Middle East Technical University} \\
{\small 06531 Ankara, Turkey} \\
{\small 2) Physics Department, Gazi University 06460 Ankara, Turkey} }

\date{}

\begin{titlepage}
\maketitle
\thispagestyle{empty}

\begin{abstract}
\baselineskip  0.7cm

We calculate the transition formfactors for the
$B\to K\ell^+\ell^-$ decay in the framework of the light
cone QCD sum rules. The invariant dilepton mass distribution
and the final lepton longitudinal polarization asymmetry
are investigated. The comparison analysis of our
results with traditional sum rules method predictions on
the formfactors is performed.
\end{abstract}

\vspace{1cm}
\end{titlepage}

\section{Introduction}

Experimental observation \cite{R1} of the inclusive
and exclusive radiative decays $B \rar X_s \gamma$ and $B \rar K \gamma$
stimulated the study of rare $B$ decays on a new footing.  These Flavor
Changing Neutral Current (FCNC) $b \rar s$ transitions in the 
Standard Model (SM) do not
occur at the tree level and appear only at the loop  level.  Therefore the
study of these rare $B$-meson decays can provide a means of testing the
detailed structure of the SM at the loop level. These decays are also very
useful for extracting the values of the Cabibbo-Kobayashi-Maskawa (CKM)
matrix elements \cite{R2}, as well as for establishing new  physics beyond
the SM \cite{R3}.

Currently, the main interest on rare $B$-meson decays is focused on
decays for which the SM predicts large branching ratios  and can be potentially
measurable in the near future. The rare \tep and
$B \rar K^* \ell^+ \ell^-$
decays are such decays. The experimental situation for these decays is very
promising \cite{R4}, with $e^+ e^-$ and hadron colliders focusing only on
the observation of exclusive modes with $\ell = e,~\mu$ and $\tau$ final states,
respectively.  At quark level the process $b \rar s \ell^+ \ell^-$ takes place
via electromagnetic and Z penguin and W box diagrams and are described by
three independent Wilson coefficients $C_7,~C_9$ and $C_{10}$.
Investigations
allow us to study different structures, described by the above mentioned
Wilson coefficients. In the SM, the measurement of the forward-backward
asymmetry and invariant dilepton mass distribution in
$b \rar q \ell^+ \ell^-~(q = s,~d)$
provide information on the short distance contributions dominated by the
top quark loops and are essential in separating the short distance FCNC
process from the contributing long distance effects \cite{R5} and also are very
sensitive to the contributions from new physics \cite{R6}. Recently it has
been emphasized by Hewett \cite{R7} that the longitudinal
lepton polarization, which is another parity violating observable, is also
an important asymmetry and that the lepton polarization in $b \rar s \ell^+
\ell^-$
will be measurable with the high statistics available at the B-factories
currently under construction.
However,
in calculating the Branching ratios and other observables in hadron level,
i.e. for \tep decay, we have the problem
of computing the matrix element of the
effective Hamiltonian, ${\cal H}_{eff}$, between the states $B$ and $K$.
This problem is related to the non-perturbative sector of QCD.

These matrix elements, in the framework of different approaches such as
chiral theory \cite{R8}, three point QCD sum rules method \cite{R9},
relativistic quark model by the light-front formalism \cite{R10,R11}, have
been investigated.
The aim of this work is the calculation of these matrix elements in light
cone QCD sum rules method and to study the lepton polarization asymmetry for
the exclusive \tep decays.

The effective Hamiltonian for the $b \rar s \ell^+ \ell^-$ decay, including
QCD corrections [12--14] can be written as
\bea
{\cal H}_{eff} = \frac{4 G_F}{\sqrt 2} V_{tb} V^*_{ts} \sum_{i=1}^{10}
C_i( \mu ) O_i( \mu )~,
\eea
which is evolved from the electroweak scale down to $\mu \sim m_b$ by the
renormalization group equations. Here $V_{ij}$ represent the relevant CKM
matrix elements, and $O_i$ are a complete set of renormalized dimension 5
and 6 operators involving light fields which govern the $b \rar s$
transitions and $C_i ( \mu )$ are the Wilson coefficients for the corresponding
operators. The explicit forms of  $C_i ( \mu )$ and $O_i ( \mu )$ can be
found in [12--14]. For $b \rar s \ell^+ \ell^-$ decay, this
effective Hamiltonian leads to the matrix element
\bea
{\cal M} = \frac{G_F \alpha}{\sqrt 2 \pi} V_{tb} V^*_{ts} \left[ C_9^{eff}
\bar s \gamma_\mu L b \, \bar \ell \gamma^\mu \ell + C_{10} \bar s \gamma_\mu L b \,
\bar \ell \gamma^\mu \gamma_5 \ell - 2 \frac{C_7}{p^2}\bar s i \sigma_{\mu \nu}p^\nu
(m_b R + m_s L) b  \, \bar \ell \gamma^\mu \ell \right]~,
\eea
where $p^2$ is the invariant dilepton mass, and
$L(R)=\left[1-(+)\gamma_5\right]/2$ are the projection operators.
  The coefficient $C_9^{eff} (
\mu,~p^2) \equiv C_9( \mu) + Y ( \mu,~p^2)$,
where the function $Y$ contains the
contributions from the one loop matrix element of the four-quark
operators, can be found in [12--14]. Note that the function $Y
( \mu,~p^2)$ contains both real and imaginary parts (the imaginary part
arises when the c-quark in the loop is on the mass shell).

The \tep decay also receives large long
distance contributions from the
cascade process $B \rar K J/\psi(\psi^\prime)\rar K \ell^+ \ell^-$.
These contributions are
taken into account by introducing a Breit-Wigner form of the resonance
propagator and this procedure leads to an additional contribution to
$C_9^{eff}$ of the form \cite{R15}
\bea
-\, \frac{3 \pi}{\alpha^2} \sum_{V=
  J/\psi,~ \psi^\prime} \frac{m_V \Gamma(V \rar \ell^+
\ell^-)}{(q^2 - m_V^2) - i m_V \Gamma_V}~.
\nnb
\eea

As we noted earlier, in order to calculate the branching ratios for the exclusive
\tep  decays, the matrix elements
$\la K \ve \bar s \gamma_\mu (1- \gamma_5) q \ve B \ra$  and
$\la K \ve \bar s i \sigma_{\mu \nu} p^{\nu} (1+\gamma_5) q \ve B \ra$
must be calculated.
These matrix elements can be parametrized in terms of the formfactors as follows
(see also \cite{R9}):
\bea
\la K ( q )  \vel \bar s \gamma_\mu ( 1-\gamma_5 ) q \ver B ( p+q ) \ra
 &=& 2q_\mu f^+(p^2)+\left[ f^+(p^2)+f^-(p^2) \right] p_\mu~, \\
\la K ( q ) \vel \bar s i \sigma_{\mu\nu} p^\nu
( 1 + \gamma_5 ) q \ver B (  p+q ) \ra
 &=& \left[ P_\mu p^2-p_\mu(Pp)\right] \frac{f_T(p^2)}{m_B+m_K}~,
\eea
where $p+q$ and $q$ are the momentum of
$B$ and  $K$ and  $P_\mu = \ga p + 2q \dr_\mu$.

\section{Sum rules for the $B\to K$ transition formfactors}

According to the QCD sum rules ideology, in order
to calculate these formfactors, we start by considering the
representation of a suitable correlator function in
hadron and quark--gluon languages. For this purpose,
we consider the following matrix elements of the
T--product of two currents between the vacuum
state and the $K$--meson:
\bea
\Pi_\mu^{(1)}(p,q) &=& i\int d^4x~e^{ipx}
   \la K(q) \ve T\{\bar s(x)\gamma_\mu(1-\gamma_5)b(x)
  \, \bar b(0)i\gamma_5q(0)\} \ve  0 \ra~, \\
\Pi_\mu^{(2)}(p,q)&=&i\int d^4x~e^{ipx}
   \la K(q)\ve T\{ \bar s (x)i\sigma_{\mu\nu}p^\nu(1+\gamma_5)b(x)
  \, \bar b (0)i\gamma_5q(0)\} \ve  0 \ra~,
\eea
where $q$ is $K$--meson momentum and $p$ is the transfer momentum.

The hadronic (physical) part of eqs.(5) and (6) is obtained by
inserting a complete set of states including the $B$--meson
ground state, and higher states with $B$--meson quantum number:

\newpage

\bea
\lefteqn{
\Pi_\mu^{(1)}(p,q) = } \nnb \\
&& = \la K(q) \ve \bar s(x) \gamma_\mu(1-\gamma_5)b(x)
\ve B(p+q)\ra \frac{1}{m_B^2-(p+q)^2}
\la B(p+q) \ve \bar b(0) i\gamma_5 q(0) \ve 0 \ra + \nnb \\
&& \; \; \; \; +\sum_h 
\la K(q) \ve \bar s(x) \gamma_\mu(1-\gamma_5)b(x) \ve h(p+q) \ra
\frac{1}{m_B^2-(p+q)^2}
\la h(p+q) \ve \bar b(0) i\gamma_5 q(0) \ve 0 \ra \nnb \\
&&=F_1(p^2,(p+q)^2) q_\mu + F_2(p^2,(p+q)^2) p_\mu~, \\ \nnb \\ \nnb  \\
\lefteqn{
\Pi_\mu^{(2)}(p,q) = } \nnb \\
&& = \la K(q) \ve \bar s(x)
i\sigma_{\mu\nu} p^\nu (1+\gamma_5) b(x) \ve B(p+q) \ra
\frac{1}{m_B^2-(p+q)^2}
\la B(p+q) \ve \bar b(0) i \gamma_5 q(0) \ve 0 \ra + \nnb \\
&& \; \; \; \; +  \sum_h 
\la  K(q) \ve \bar s(x)
i\sigma_{\mu \nu}p^\nu (1+\gamma_5) b(x) \ve  h(p+q) \ra
\frac{1}{m_B^2-(p+q)^2}
\la h(p+q) \ve \bar b(0) i\gamma_5 q(0) \ve 0 \ra  \nnb\\
&&= F_3((p+q)^2,p^2) \left[P_\mu p^2-p_\mu(Pp) \right]~,
\eea
where $P_\mu=(p+2q)_\mu$. Then, for the invariant amplitudes
$F_i$, one can write a general dispersion relation in  
the $B$ meson momentum squared, $(p+q)^2$, as:
\beq
F_i(p^2,(p+q)^2))=\int_{m_b^2}^\infty\frac{\rho_i(p^2,s)ds}
                  {s-(p+q)^2}~,
\eeq
where the spectral densities are given by
\bea
\rho_1(p^2,s) &=& \delta(s-m_B^2)2f^+(p^2)
\frac{m_B^2f_B}{m_b}+\rho^h_1(p^2,s)~, \\ \nnb \\
\rho_2(p^2,s) &=& \delta(s-m_B^2)\left[f^+(p^2)+f^-(p^2) \right]
\frac{m_B^2f_B}{m_b}+\rho_2^h(p^2,s)~, \\ \nnb \\
\rho_3 (p^2,s) &=& \delta(s-m_B^2)\frac{f_T(p^2)}{m_B+m_K} ~
\frac{m_B^2f_B}{m_b} + \rho_3^h(p^2,s)~.
\eea
The first terms in eqs.(10), (11) and (12) represent the ground state
$B$--meson contribution and follow from eqs.(5) and (6) by inserting
the matrix elements given in eqs.(3), (4) and by replacing
\bea
\la B \ve \bar  b i\gamma_5 q \ve 0 \ra =\frac{f_Bm_B^2}{m_b}~,
\eea
in which we neglect the mass of the light quarks.
In eqs.(10)-(12), $\rho_i^h(p^2,s)$ represent the spectral density of
the higher resonances  and of the continuum of states.
In accordance with the QCD sum rules method we invoke
the quark--hadron duality prescription and replace the spectral density 
$\rho^h_i$ by
\bea
\rho_i^h(p^2,s)=\frac{1}{\pi}\mbox{Im}F_i^{QCD}(p^2,s)\Theta(s-s_0)~,
\eea
where Im$F_i^{QCD}(p^2,s)$ is obtained from the  imaginary part of the 
correlator functions calculated in QCD starting from some threshold $s_0$.
In order to suppress the higher states and continuum contributions,
we follow the standard procedure in the  QCD sum rules method and apply
Borel transformation $\hat B$ on the variable $(p+q)^2$ to the dispersion
integral to get,
\bea
F_i(p^2,M^2)&=&\hat B F_i(p^2,(p+q)^2) \nnb \\
&=&\int_{m_b^2}^\infty \rho_i(p^2,s)e^{-s/M^2}ds~. 
\eea
Using eqs.(10)-(12) and (14) we get
\bea
F_1(p^2,M^2) &=& 2f^+(p^2) \frac{m_B^2f_B}{m_b}e^{-m_B^2/M^2}+
    \frac{1}{\pi}\int_{s_0}^\infty \mbox{Im}F_1^{QCD}(p^2,s) e^{-s/M^2}ds~, \\ \nnb \\
F_2(p^2,M^2) &=& \left[ f^+(p^2)+
f^-(p^2)\right] \frac{m_B^2f_B}{m_b} e^{-m_B^2/M^2}+ \nnb \\ 
&&+\, \frac{1}{\pi}\int_{s_0}^\infty \mbox{Im}F_2^{QCD}(p^2,s)e^{-s/M^2}ds~, \\
\nnb \\
F_3(p^2,M^2) &=& \frac{f_T(p^2)}{M_B+m_K}\frac{m_B^2f_B}{m_b}
         e^{-m_B^2/M^2} +\frac{1}{\pi}\int_{s_0}^\infty\mbox{Im}
                       F_3^{QCD}(p^2,s)e^{-s/M^2}ds~.
\eea
The main problem is then the calculation of the correlator
functions (5) and (6) in QCD. After applying the Borel transformation, the result 
can be written in the following form:
\beq
F_i(p^2,M^2)=\frac{1}{\pi}\int_{m_b^2}^\infty\mbox{Im}
                       F_i^{QCD}(p^2,s)e^{-s/M^2}ds~.
\eeq
Equating the  $F_i$  in eq.(19) to the corresponding
$F_i$ in eqs.(16)-(18), we arrive at the sum rules for
the formfactors, which describe the $B \rar K$ transition:
\bea
f^+(p^2) &=&\frac{m_b}{2\pi f_Bm_B^2}\int_{m_b^2}^{s_0}\mbox{Im}
   F_1^{QCD}(p^2,s)e^{-(s-m_B^2)/M^2}ds~, \\ \nnb \\
f^+(p^2)+f^-(p^2) &=& \frac{m_b}{\pi f_Bm_B^2}\int_{m_b^2}^{s_0}\mbox{Im}
   F_2^{QCD}(p^2,s)e^{-(s-m_B^2)/M^2}ds~, \\ \nnb \\
f_T(p^2) &=& \frac{m_b}{\pi f_Bm_B^2}(m_B+m_K)\int_{m_b^2}^{s_0}\mbox{Im}
   F_3^{QCD}(p^2,s)e^{-(s-m_B^2)/M^2}ds~.
\eea
One can calculate Im$F_i^{QCD}(p^2,s)$,
in the deep Euclidean region, where both $p^2$
and $(p+q)^2$ are negative and large. The leading contribution
to the operator product expansion comes from the contraction of the
$b$--quark operators expressed in eqs.(5) and (6) to the free $b$--quark
propagator 
\bea
\la 0 \ve T\{b(x) \bar  b(0) \} \ve 0 \ra =-i\int\frac{d^4k}{(2\pi)^4}
e^{-ikx} \frac{\not\!k + m_b}{m_b^2-k^2}~. \nnb 
\eea
\newpage
Then we have
\bea
\Pi_\mu^{(1)}(p,q) &=&i \int\frac{d^4x \, d^4k}{(2\pi)^4(m_b^2-k^2)}
   e^{i(p-k)x} \times \nnb \\ \nnb \\
&& \times \la K(q) \ve \left[ -m_b \bar s \gamma_\mu(1-\gamma_5)q
      +\bar  s\gamma_\mu \not\!k(1+\gamma_5)q \right]\ve 0 \ra~, \\ \nnb \\
\Pi_\mu^{(2)}(p,q) &=& -\int\frac{d^4x \, d^4k}{(2\pi)^4(m_b^2-k^2)}
   e^{i(p-k)x} p^\nu \times \nnb \\ \nnb \\ 
&& \times  \la K(q) \ve \left[ - \bar s
         \sigma_{\mu \nu} \not\!k(1-\gamma_5)q
      +m_b \bar s\sigma_{\mu\nu}(1+\gamma_5)q \right]\ve 0 \ra~.
\eea
Note that, as can be seen from eqs.(23) and (24), the problem is reduced to the
calculation of the  matrix elements of the gauge--invariant,
nonlocal operators  sandwiched in between the vacuum and the
$K$ meson states. These matrix elements define the $K$ meson
light cone wave functions. Following [16--17], we define
the $K$--meson wave functions as:
\bea
\la K(q) \ve \bar  s(x) \gamma_\mu(1-\gamma_5)q(0) \ve 0 \ra&=&
    iq_\mu f_K\int_0^1 du ~e^{iuqx}\left[ \varphi_K(u)+x^2g_1(u)\right] - \nnb \\
    && - f_K \ga x_\mu-\frac{x^2 q_\mu}{qx} \dr \int_0^1 du~e^{iuqx}g_2(u)~.
\eea
In eq.(25) $\varphi_K(u)$ is the leading twist two,  $g_1(u)$
and $g_2(u)$ are the twist four $K$ meson wave functions, respectively.
The second matrix element of eq.(23), can be split
into two matrix elements using the identity
$\gamma_\mu\gamma_\nu=g_{\mu\nu}-i\sigma_{\mu\nu}$, and the
result can be evaluated using the twist
three wave functions defined as [16--19]:
\bea
\la K(q) \ve \bar s(x) i \gamma_5 q(0) \ve 0 \ra &=&
\frac{f_K m_K^2}{m_s + m_q}\int_0^1 du~e^{iuqx}\varphi_p(u)~, \\ \nnb \\
\la K(q) \ve \bar s(x) \sigma_{\mu \nu}(1+\gamma_5)q(0) \ve 0 \ra &=&
 i (q_\mu x_\nu-q_\nu x_\mu)
  \frac{f_K m_K^2}{6(m_s + m_q)}\int_0^1 du~e^{iuqx}\varphi_\sigma(u)~.
\eea
The matrix elements in  eq.(24) can be easily calculated using the
identities,
\bea
\sigma_{\mu\nu} &=&-\frac{i}{2}\epsilon_{\mu\nu\rho\beta}
     \sigma^{\rho\beta}\gamma_5  \nnb \\
\sigma_{\mu\nu}\gamma_\rho &=&i(\gamma_\mu g_{\rho\nu}-
    \gamma_\nu g_{\rho\mu})+\epsilon_{\mu\nu\rho\beta}
     \gamma^\beta\gamma_5~, \nnb
\eea
to express it  in terms of the wave functions in eq.(25).

Substituting the matrix elements (25--27) into the eqs.(23) and (24)
and integrating over the variables $x$ and $k$ we get the following
expressions for $F_i(p^2,(p+q)^2)$:
\bea
F_1(p^2,(p+q)^2) &=& 
m_b f_K \int_0^1 \frac{du}{\Delta}
  \left[\varphi_K(u)-\frac{8 m_b^2 \left[ g_1(u)+G_2(u)\right] }{\Delta^2}+
           \frac{2 u g_2(u)}{\Delta} \right] +  \nnb \\
&+& f_K \mu_K \int_0^1 \frac{du}{\Delta}\left[u\varphi_\rho(u)
     +\frac{1}{6}\varphi_\sigma(u) \ga 1+
                \frac{2m_b^2-2upq-2q^2u^2}{\Delta} \dr \right]~,  \\ \nnb \\
F_2(p^2,(p+q)^2) &=& m_b f_K \int_0^1 \frac{du}{\Delta}
\Bigg\{ \Bigg[ \frac{2 g_2(u)}{\Delta}+\frac{\mu_K}{m_b} 
\Bigg( \varphi_p(u) - \frac{(2pq+2q^2u)}{6\Delta}
        \varphi_\sigma(u) \Bigg) \Bigg] \Bigg\}~, \\ \nnb \\ 
F_3(p^2,(p+q)^2) &=& - f_K \int_0^1 \frac{du}{\Delta}
\left\{ \frac{1}{2} \left[ \varphi_K(u)-\frac{4 \left[g_1(u)+G_2(u)\right]}{\Delta}
 \ga 1+\frac{2m_b^2}{\Delta}\dr \right]+  \right. \nnb \\ \nnb \\
&+& \left. m_b\mu_K\frac{\varphi_\sigma(u)}{6\Delta} \right\}~,
\eea
where
$$
q^2=m_K^2,~~\Delta=m_b^2-(p+qu)^2~,~~\mu_K=\frac{m_K^2}{m_s+m_q}~,~~
       G_2(u)=-\int_0^ug_2(v)dv~.
$$

There are also
contributions to the above considered wave functions 
from multi-particle meson wave functions.
Here we consider only the operator $\bar q G q$, which gives the 
main  contribution and corresponds to the quark--antiquark--gluon
components in the kaon (for more detail see \cite{R20}).
In this approximation, the $b$--quark propagator is defined as
 (see \cite{R18}):
\bea
\Big{\la} 0 \vel T\left\{ b(x) \bar b(0)\right\}\ver 0 \Big{\ra} &=&
  i S_b^0(x)-ig_s \int \frac{d^4k}{(2\pi)^4}~ e^{-ikx} \times \nnb \\ \nnb \\
&\times&    \int_0^1 du \left[ \frac{1}{2}\frac{\not\!k+m_b}{(m_b^2-k^2)^2}
        G^{\mu\nu}(ux) \sigma_{\mu \nu} + \frac{1}{m_b^2-k^2}
        u x_\mu G^{\mu \nu}(ux) \gamma_\nu \right]~. \nnb \\ \nnb \\ 
\eea
It is clear that when we take into account the second term
in the right hand side of eq.(31), new matrix elements appear.
Substituting eq.(31) into eqs.(5) and (6), and using the identity
\bea
\gamma_\mu \gamma_\nu \sigma_{\rho\lambda}&=&
   (\sigma_{\mu\lambda}g_{\nu\rho}-\sigma_{\mu\rho}g_{\nu\lambda})
  +i(g_{\mu\lambda}g_{\nu\rho}-g_{\mu\rho}g_{\nu\lambda})-\nnb\\
&-&\epsilon_{\mu\nu\rho\lambda}\gamma_5-
   i\epsilon_{\nu\rho\lambda\alpha}
    g^{\alpha\beta}\sigma_{\mu\beta}\gamma_5~,\nnb
\eea
one can express the resulting new matrix elements in terms of the
three particle wave functions [16--20]:
\bea
\la K(q)\ve \bar s(x) g_s G_{\mu\nu}(ux)\sigma_{\alpha\beta}\gamma_5 q(0) \ve 0 \ra &=&   
if_{3K}\big[ (q_\mu q_\alpha g_{\nu\beta} - q_\nu q_\alpha g_{\mu\beta})  
- \left( q_\mu q_\beta g_{\nu\alpha}-q_\nu q_\beta
g_{\mu\alpha} \right) \big] \times \nnb \\
&\times& \int {\cal D}\alpha_i\varphi_{3K}(\alpha_i)
  e^{iqx \omega}~, \\ \nnb \\
\la K(q) \ve \bar s(x)\gamma_\mu\gamma_5 g_s
                  G_{\alpha\beta}(ux)q(0) \ve 0 \ra &=&  
f_K \left[ q_\beta \ga g_{\alpha\mu}-\frac{x_\alpha q_\mu}{qx} \dr-
     q_\alpha \ga g_{\beta\mu}-\frac{x_\beta q_\mu}{qx} \dr \right] \times \nnb \\
&\times&     \int{\cal D}\alpha_i\varphi_\perp(\alpha_i)
     e^{iqx \omega} + \nnb\\
&+& f_K \frac{q_\mu}{qx}\ga q_\alpha x_\beta-q_\beta x_\alpha\dr
       \int{\cal D}\alpha_i\varphi_\parallel(\alpha_i)
     e^{iqx \omega}~,  \\ \nnb  \\
\la K(q) \ve \bar s(x)\gamma_\mu g_s
                           \tilde G_{\alpha\beta}(ux)q(0) \ve 0 \ra &=&  
i f_K \left[ q_\beta \ga g_{\alpha\mu}-\frac{x_\alpha q_\mu}{qx} \dr -
     q_\alpha \ga g_{\beta\mu}-\frac{x_\beta q_\mu}{qx} \dr \right] \times \nnb \\
&\times& \int{\cal D}\alpha_i\tilde\varphi_\perp(\alpha_i)
     e^{iq x \omega} + \nnb  \\
&+& i f_K \frac{q_\mu}{qx}(q_\alpha x_\beta-q_\beta x_\alpha)
       \int{\cal D}\alpha_i\tilde\varphi_\parallel(\alpha_i)
     e^{iqx \omega}~,
\eea
where $\tilde G_{\alpha\beta}=\frac{1}{2}\epsilon_{\alpha\beta\rho\lambda}
G^{\rho\lambda}$, ${\cal D}\alpha_i=d\alpha_1 d\alpha_2 d\alpha_3
\delta(1-\alpha_1-\alpha_2-\alpha_3)$, and $\omega = \alpha_1 + u \alpha_3$.
Here $\varphi_{3K}(\alpha_i)$
is a twist three wave function and the remaining functions $\varphi_\perp$,
$\varphi_\parallel$, $\tilde\varphi_\perp$ and $\tilde\varphi_\parallel$
are all twist four wave functions.

When we substitute eqs.(32)--(34) into  eqs.(5) and (6), perform integration
over $x$ and $k$, add (28), (29), (30), apply the Borel
transformation over $(p+q)^2$ and equate the obtained  results to (16),
(17) and (18),  
we  get the following sum rules for the $B \rar K$ transition
formfactors:
\bea
f^+(p^2)&=&\frac{m_bf_K}{2f_Bm_B^2}\,e^{\frac{m_B^2}{M^2}}
\left\{ \int_\delta^1~exp\left(-\frac{m_b^2-p^2(1-u)+q^2u(1-u)}{uM^2}
\right)\frac{du}{u} \times \right.\nnb \\
&\times& \left.  \Bigg[ m_b \Bigg( \varphi_K(u) -\frac{8m_b^2 \left[ g_1(u)+G_2(u) \right]}{2u^2M^4}+
\frac{2g_2(u)}{M^2} \Bigg) + \right. \nnb \\
&+& \left.  \mu_K
\left( u\varphi_\rho(u) + \frac{\varphi_\sigma(u)}{6}
                    \ga 2+\frac{m_b^2+p^2-q^2u^2}{uM^2}\dr \right)\Bigg] +\right.  \nnb\\
&+& \left.  f_{3K}~\int_0^1 du \int{\cal D}\alpha_i \, \theta
(\omega-\delta)\,  exp \left( -\frac{m_b^2-p^2(1-\omega)+q^2\omega
            \left(1-\omega \right)}{ \omega M^2} \right)  \times
\right. \nnb\\
&\times&\left.  \left[ \frac{(2u-1)\varphi_{3K}}{f_K}\frac{3q^2}
          {\omega M^2} +\frac{2u \varphi_{3K}}{f_K} \left(-\frac{1}{\omega^2}
          + \frac{m_b^2-p^2-q^2\omega^2}
                 {\omega^3M^2}\right) + \right.
\right. \nnb \\
&+&\left. \left. \frac{m_b}{f_{3K}} \left(\frac{2 \varphi_\perp-\varphi_\parallel+2\tilde\varphi_\perp-
   \tilde\varphi_\parallel}{\omega^2 M^2}
   + \frac {2 q^2\alpha_3
    (\Psi_\perp+\Psi_\parallel+\tilde\Psi_\perp+\tilde\Psi_\parallel)}
{\omega^2 M^4}\right) \right] \right\}~, \\ \nnb \\ \nnb \\
f^+(p^2)+f^-(p^2)&=&\frac{m_bf_K}{f_Bm_B^2}\,e^{\frac{m_B^2}{M^2}}
\left\{\int_\delta^1~exp \left(-\frac{m_b^2-p^2(1-u)+q^2u(1-u)}{u M^2}
       \right) \frac{du}{u}\times \right.\nnb\\
&\times& \left.  \left[2m_b\frac{g_2(u)}{u M^2}+\mu_K \left( \varphi_\rho(u)+\frac{\varphi_\sigma(u)}{6u}-
\frac{\varphi_\sigma(u)}{6u^2M^2}(m_b^2-p^2+q^2u^2) \right) \right]+ \right.\nnb\\
&+&\left.  \int_0^1du\int{\cal D}\alpha_i \, \theta(\omega-\delta)
 \, exp \left(-\frac{m_b^2-p^2(1-\omega)+
        q^2\omega(1-\omega)}
                 {\omega M^2} \right)\frac{q^2}{\omega^2 M^2} \times \right. \nnb\\
&\times&\left.  \left[\frac{f_{3 K}(2u-3)\varphi_{3K}}{f_K}+2m_b\alpha_3\frac{\Psi_\perp+\Psi_\parallel
        +\tilde\Psi_\perp+\tilde\Psi_\parallel}
                         {\omega M^2}\right] \right\}~, 
\eea
\newpage
\bea
f_T(p^2)&=& \frac{m_b(m_B+m_K)f_K}{f_Bm_B^2}\,e^{\frac{m_B^2}{M^2}}
\left\{ \int_\delta^1 du~exp \left(-\frac{m_b^2-p^2(1-u)+
    q^2u(1-u)}{uM^2} \right) \times \right.\nnb\\
&\times& \left.  \left[-\mu_K\frac{m_b\varphi_\sigma(u)}{6u^2M^2}-\frac{1}{2}
    \frac{\varphi_K(u)}{u}+
2 \left(\frac{m_b^2}{u M^2} + 1 \right) \frac{ \left[ g_1(u)+G_2(u)
\right]}{u^2 M^2}\right] + \right. \nnb\\
&+& \left. \int_0^1 du \int{\cal D}\alpha_i \, \theta
(\omega-\delta) \, exp\left(-\frac{m_b^2-p^2(1-\omega)+q^2\omega
(1-\omega)}{\omega M^2}\right) \times \right. \nnb \\
&\times& \left.  \left[ \frac{u  \left(\varphi_\parallel-2\tilde\varphi_\perp \right)}
   {\omega^2 M^2} + \frac{\varphi_\parallel + \tilde \varphi_\parallel - 2
\varphi_\perp - 2 \tilde \varphi_\perp}{\omega^2 M^2}\right] \right\}~, 
\eea
where $\delta = \frac{m_b^2 - p^2}{s_0 - p^2}$.
The functions $\Psi_\perp(\tilde\Psi_\perp)$,
$\Psi_\parallel(\tilde\Psi_\parallel)$ in eqs.(35)--(37) are defined in the
following way
\begin{eqnarray}
&&\Psi_\perp(\tilde\Psi_\perp)=
           -\int_0^u\varphi_\perp(v)(\tilde\varphi_\perp(v))dv~,\nnb\\
&&\Psi_\parallel(\tilde\Psi_\parallel)=
   -\int_0^u\varphi_\parallel(v)(\tilde\varphi_\parallel(v))dv~.\nnb
\end{eqnarray}
Note that the formfactor $f^+(p^2)$  is
investigated in \cite{R18} and \cite{R19} (In \cite{R18}, it is necessary to make
the simple replacement $\pi\to K$), in light cone sum rules. Our results coincide with
theirs if we let $q^2=0$.

At the end of this section we calculate the differential decay rate
with the longitudinal polarization of the final leptons and we obtain
\bea
\frac{d \Gamma}{d p^2} &=&
\frac{G^2 \alpha^2}{2^{12} \pi^5} \,
\frac{\vel V_{tb} V_{ts}^* \ver ^2 v \sqrt{\lambda}}{m_B}
\Bigg\{m_B^2 (2 m_\ell^2 + m_B^2 s)\left( \vel A \ver^2 + \vel C \ver^2 \right)
\frac{\lambda}{3 s} + \nnb \\
&+& 8 m_B^2 m_\ell^2 \left[  r \vel C \ver^2 + s \vel D
\ver^2  + Re( C^* D) (1-r-s) \right] +\nnb \\
&-&  \frac{2}{3} Re ( A^* C) m_B^4 \xi v
\lambda \Bigg\} ~,
\eea
where $\lambda=1+r^2+s^2-2r-2s-2sr,~~r=m_K^2/m_B^2,~~
s=p^2/m_B^2,~~\xi$ is the longitudinal polarization of the
final lepton, $m_\ell$ and $v=\sqrt{1-\frac{4 m_\ell^2}{p^2}}$ are
its mass and velocity, respectively. In eq.(38) $A$, $C$
and $D$ are defined as follows:
\bea
&&A= 2C_9^{eff}f^+- C_7\frac{4m_bf_T}{m_B+m_K}~,\nnb\\
&&C= 2 C_{10}f^+~, \\
&&D= C_{10}(f^+ + f^-)~.\nnb
\eea
For the dileptonic decays of the $B$ mesons, the longitudinal
polarization asymmetry $P_L$ of the final state $\ell$,
is defined by
\begin{equation}
P_L(p^2)=\frac{\displaystyle{\frac{d\Gamma}{dp^2}(\xi=-1)-
                  \frac{d\Gamma}{dp^2}(\xi=1)}}
                {\displaystyle{\frac{d\Gamma}{dp^2}(\xi=-1)+
                  \frac{d\Gamma}{dp^2}(\xi=1)}}~.
\end{equation}
where $\xi=-1(+1)$ corresponds to the left (right) handed
lepton. If in eq.(40), we let $m_\ell=0$, our results coincide with
the results in \cite{R21} and if $m_\ell\neq0$, they coincide with
the ones in \cite{R11}.

\section{Numerical analysis}

The main input parameters in the sum rules
(35)--(37) are the kaon wave functions on the light cone.
For kaon wave functions
we use the results of ref. \cite{R16,R17,R19}:
\begin{eqnarray}
&&\varphi_K=6u(1-u) \left\{1+0.52\left[5(2u-1)^2-1\right]+0.34\left[21(2u-1)^4- 14
(2u-1)^2+1\right]\right\}~, \nnb \\ 
&&\varphi_p\simeq 1~,\nonumber\\
&&\varphi_\sigma\simeq 6u(1-u)~, \nonumber\\
&&g_1(u)\simeq\frac{5}{2}\delta^2u^2(1-u)^2 ~, \nonumber\\
&&g_2(u)\simeq \frac{10}{3}\delta^2u(1-u)(2u-1)~, \nonumber\\
&&\varphi_{3K}(\alpha_i)\simeq 360 \alpha_1\alpha_2\alpha_3^2~,\nonumber\\
&&\varphi_{\perp}(\alpha_i)\simeq 10\delta^2
       (\alpha_1-\alpha_2)\alpha_3^2~,\nonumber\\
&&\varphi_{\parallel}(\alpha_i)\simeq
     120\delta^2\epsilon(\alpha_1-\alpha_2)
            \alpha_1\alpha_2\alpha_3 ~,  \nonumber\\
&&\tilde\varphi_{\perp}(\alpha_i)\simeq 10\delta^2
       \alpha_3^2(1-\alpha_3)~,\nonumber\\
&&\tilde\varphi_{\parallel}(\alpha_i)\simeq
     -40\delta^2\alpha_1\alpha_2\alpha_3~,   \nonumber
\end{eqnarray}
here $\delta^2(\mu_b)\simeq 0.17~ GeV^2$ at $\mu_b\simeq
\sqrt{m_B^2-m_b^2}=2.4~ GeV^2$, which follows from the QCD sum rules analysis
(for more detail see \cite{R18}), $\epsilon(\mu_b)\simeq 0.36$.
In order to estimate  $\mu_K$, we use the PCAC relation for the pseudo
Goldstone bosons
$$
\frac{\mu_K}{\mu_\pi}=\frac{\left(\la qq \ra +\la \overline ss \ra \right)f_K^2 }
         {2 \la qq \ra f_\pi^2}\simeq0.62~~~~(q=u \mbox{ or } d)~.
$$
Here we use $f_\pi\simeq133~MeV,~f_K\simeq 160~MeV$ and we assume 
 $ \la \overline q q \ra \div \la \overline s s \ra=1\div 0.8$, which follows from QCD
sum rules for strange hadrons \cite{R22}. As a result we have
$$
\mu_K\simeq 1 ~GeV~\mbox{when}~ \mu_\pi=\frac{m^2_\pi}{m_u+m_d}
    \simeq 1.6~ GeV~.
$$
For the values of the other parameters, we choose:
$f_B\simeq 0.14~GeV$, which is obtained
from 2--point QCD sum rules analysis \cite{R18,R23},
$m_b\simeq 4.7~ GeV$ and $s_0\simeq 35~GeV^2$,
and $\left|V_{tb}V^*_{ts} \right|\simeq 0.045$.

Before giving numerical results on the formfactors,
we must first determine the region for the Borel mass
parameter $M^2$, for which the sum rules yields reliable results.
The lower limit of this region is determined by the requirement
that, the terms proportional to $M^{-2n}(n>1)$ remain subdominant.
The upper limit of $M^2$ is determined by requiring the
higher resonance and continuum contributions to be less than
$\sim 30\%$ of the total result. Our numerical results show that both
requirements are satisfied in the region  $8~GeV^2\leq M^2\leq 16~GeV^2$
and for the numerical analysis we use $M^2=10~ GeV^2$. When $p^2$
approaches the region $m_b^2-O(1~GeV^2)$ a breakdown of the
stability is expected, similar to the $B \rar \pi$ case (see \cite{R18,R19}).

In fig.1, we present the $p^2$ dependence of the formfactors
$f^+(p^2)$, $f^-(p^2)$ and $f_T(p^2)$. At zero momentum transfer,
the QCD prediction for the formfactors are
\begin{eqnarray}
&&f^+(p^2=0)=0.29~, \nnb\\
&&f^-(p^2=0)=-0.21~,\nnb\\
&&f_T(p^2=0)=-0.31~.\nnb
\end{eqnarray}

In fig.2a(b) we present the $p^2$ dependence of the differential 
Branching ratios
for the $B\to K\mu^+\mu^-$ and ($B\to K\tau^+\tau^-$) decay
with and without long distance effects.
In both cases summation over the final lepton polarization is performed.
Note that $p^2$ dependence of the differential branching ratio 
$B \rar K \tau^+ \tau^-$ is analysed in \cite{R11} and \cite{R24}
using the light front formalism and heavy meson chiral theory,
respectively. In \cite{R11} both short and long distance contributions 
are considered while in \cite{R24} only the short distance
contributions are taken into account and in both works the obtained spectrum is
fully symmetric while in the present work the spectrum we obtain seems
to be slightly asymmetric as a result of 
a highly asymmetric resonance-type behaviour due to the nonperturbative
contributions.  
Performing the integration over $p^2$ in eq.(38) and using
the values of the life times $\tau_{B_d} = (1.56 \pm 0.06) \times 10^{-12}$
s,  $\tau_{B_u} = (1.62 \pm 0.06) \times 10^{-12}$ s \cite{R25}, for the branching
ratios, including only the short distance contributions,
we get:
\bea
B(B_d \rar K^0 \mu^+ \mu^-)     &=& (3.1 \pm 0.9) \times 10^{-7}~ , \nnb \\
B(B_d \rar K^0 \tau^+ \tau^-)   &=& (1.7 \pm 0.4) \times 10^{-7}~ ,\nnb \\
B(B_u \rar K^+ \mu^+ \mu^-)     &=& (3.2 \pm 0.8) \times 10^{-7}~ ,\nnb \\
B(B_u \rar K^+ \tau^+ \tau^-)   &=& (1.77\pm 0.40) \times 10^{-7}~ ,\nnb
\eea
where theoretical and experimental errors have been added quadratically.
 
In fig.3 we display the lepton longitudinal polarization asymmetry
$P_L$ as a function of $p^2$ for the $B\to K\mu^+\mu^-$ and
$B\to K\tau^+\tau^-$ decays, at $m_t=176~GeV$, with and without the
long distance contributions. From this figure one can see that
$P_L$ vanishes at the threshold due to the kinematical factor
$v$ and the value of $P_L$ for the $B\to K\mu^+\mu^-$ decay varies in the
region ($ 0\div -0.7 $) and ($ 0 \div -0.1$) for the $B\to K\tau^+\tau^-$
decay, when long distance effects are excluded.

\section{Conclusion}
In this work, we  calculate the transition formfactors for the exclusive
$B \rar K \ell^+ \ell^-$($\ell=\mu,~\tau$) decay in the framework of the light cone
QCD sum rules,
and investigate the longitudinal
polarization asymmetries of the muon and tau in this decay. 
From a comparison of our results with the traditional QCD sum rule
predictions (see \cite{R9}), we observe  that the behaviour of the
formfactors are similar and the value of $f^+$ in both approaches
coincides, while $f_T$ differs two times than that of \cite{R9} at $p^2=0$.
It is important to note that for a more refined analysis, it is
necessary to take into account the SU(3) breaking effects:
the differences between $f_K$, $\mu_K$ and $f_\pi$, $\mu_\pi$ and
the differences of pion and kaon wave functions. These SU(3)
breaking terms  can lead to 
differences between $B\to\pi$ and $B\to K$
formfactors. But we expect that these 
effects can change the results about 15--20$\%$ and this lies
at the accuracy level of the sum rules method.

Few words about the possibility of the experimental observation of this
decay are in order. Experimentally, to observe an asymmetry $P_L$ of a decay
with the branching ratio $B$ at the $n\sigma$ level, the required number of
events is $N = \frac{n^2}{B\,P_L^2}$ (see \cite{R11}). For example, to
observe the $\tau$ lepton polarization at the exclusive channel $B \rar K
\tau^+ \tau^-$ at the $3\sigma$ level, one
needs at least $N \simeq 5 \times 10^9~B\, \bar B$ decays.
Since in the future $B$-factories, it is expected that
$\sim~10^9$ $B$-mesons would be created per year,
it is possible to measure the longitudinal polarization asymmetry of the
$\tau$ lepton.

\newpage

\section*{Figure Captions}
{\bf 1.} The $p^2$ dependence of the formfactors
$f^+(p^2),~f^-(p^2)$ and $f_T(p^2)$. \\ \\
{\bf 2.}\, a) Invariant mass squared distribution of the lepton pair for the
decay $B \rar K \mu^+ \mu^-$.\\ 
\indent b) The same as in a), but for the decay \tept. \\
\indent Here and in all of the following figures the solid line corresponds to the 
short distance \indent contributions only and the 
dashed line to the sum of both short and long distance contri-\\
\indent butions.\\ \\
{\bf 3.}\, a) The longitudinal polarization asymmetry $P_L$ for the 
\tepm
decay. \\
\indent b) The same as in a), but for the \tept decay.
\\

\begin{figure}
\vspace{30.2cm}

    \includegraphics{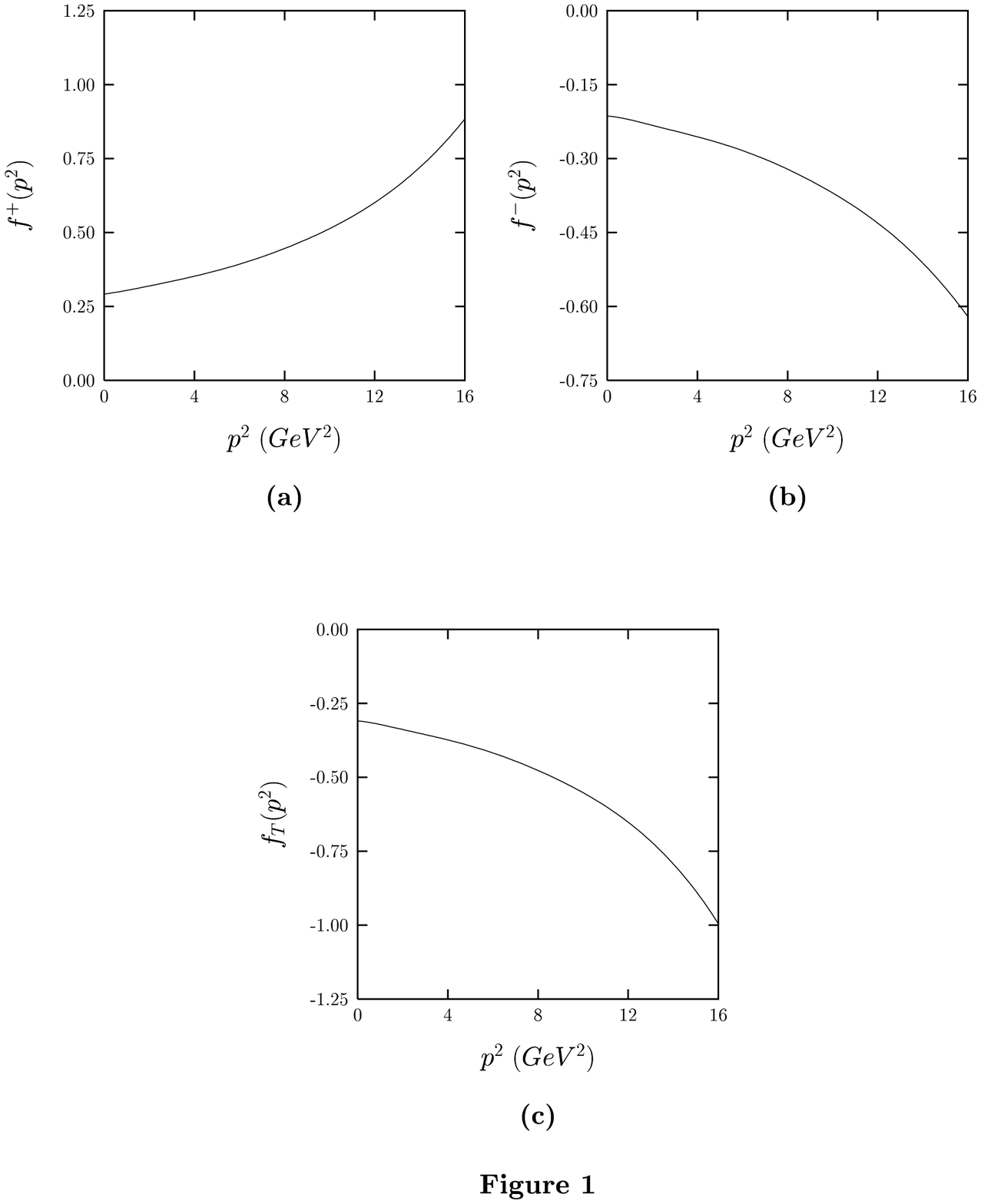}
    \vspace{-13.5cm}
\end{figure}

\begin{figure}
\vspace{30.0cm}
    \includegraphics{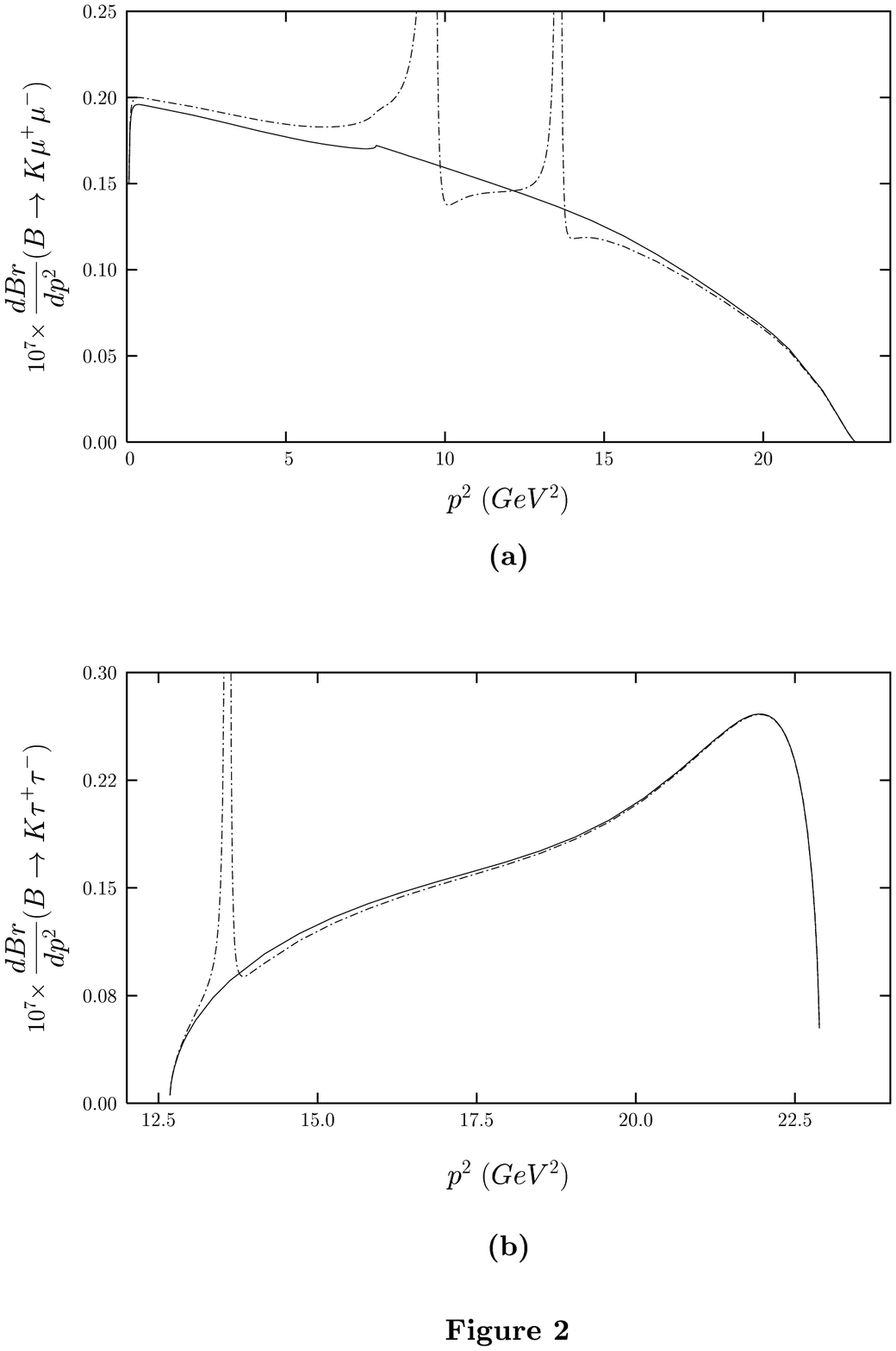}
    \vspace{-15.0cm}
\end{figure}

\begin{figure}
\vspace{30.0cm}
    \includegraphics{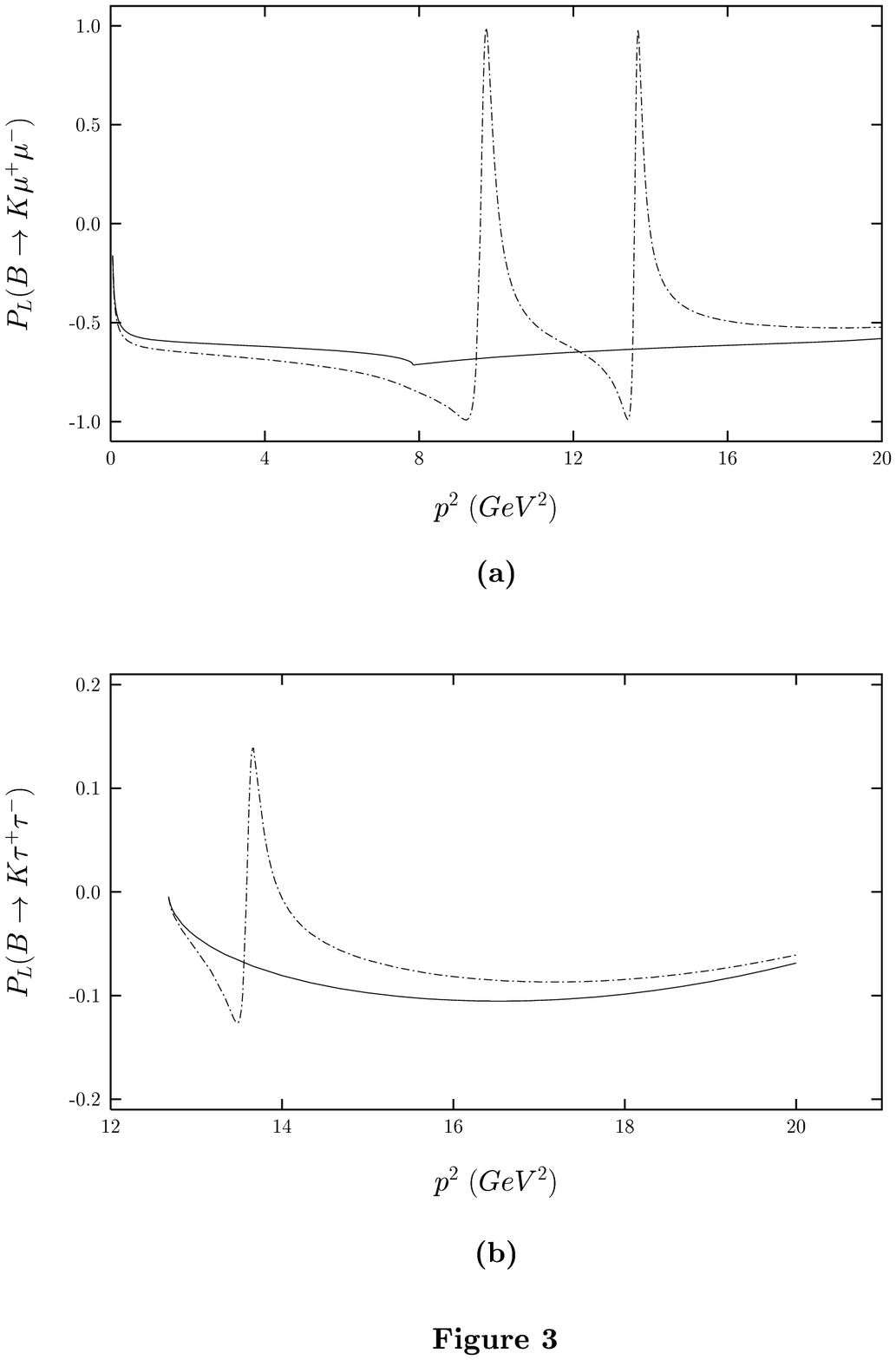}
    \vspace{-15.0cm}
\end{figure}

\end{document}